\documentclass[a4paper,doc,12pt, natbib]{apa6}

\usepackage[english]{babel}
\usepackage[utf8x]{inputenc}
\usepackage{amsmath}
\usepackage{setspace}
\usepackage{graphicx}
\usepackage[colorinlistoftodos]{todonotes}
\newcommand{\red}[1]{{\textcolor{black}{#1}}}
\usepackage{geometry}
\usepackage[most]{tcolorbox}

\usepackage{amsfonts}
\usepackage{amssymb}
\usepackage{listings}
\makeatletter
\def\Let@{\def\\{\notag\math@cr}}
\makeatother
\usepackage{amsmath,amssymb}

\usepackage{subcaption}
\usepackage{tabularx} %
\newcolumntype{L}[1]{>{\raggedright\arraybackslash}p{#1}}
\newcolumntype{C}[1]{>{\centering\arraybackslash}p{#1}}
\newcolumntype{R}[1]{>{\raggedleft\arraybackslash}p{#1}}
\usepackage{threeparttable} %
\usepackage{siunitx} %
\usepackage{pdflscape} %
\usepackage{hhline} %
\usepackage{tabularx} %
\usepackage{booktabs} %
\usepackage{color}

\usepackage[
singlelinecheck=false 
]{caption}

\title{
The co-evolution of emotional well-being with weak and strong friendship ties}%
\shorttitle{Co-evolution of well-being and friendship}
\author{Timon Elmer, Zsófia Boda, \& Christoph Stadtfeld}
\affiliation{Chair of Social Networks, ETH Zürich, Switzerland \\ 
\footnote{cite as: \\ Elmer, T., Boda, Z., \& Stadtfeld, C. (in press). The co-evolution of emotional well-being with weak and strong friendship ties. \textit{Network Science}.}
}

\abstract{Social ties are strongly related to well-being. But what characterizes this relationship? This study investigates social mechanisms explaining how social ties affect well-being through social integration and social influence, and how well-being affects social ties through social selection.
We hypothesize that highly integrated individuals -- those with more extensive and dense friendship networks -- report higher emotional well-being than others. Moreover, emotional well-being should be influenced by the well-being of close friends. Finally, well-being should affect friendship selection when individuals prefer others with higher levels of well-being, and others whose well-being is similar to theirs. We test our hypotheses using longitudinal social network and well-being data of 117 individuals living in a graduate housing community. The application of a novel extension of Stochastic Actor-Oriented Models for \textit{ordered} networks (ordered SAOMs) allows us to detail and test our hypotheses for weak- and strong-tied friendship networks simultaneously. Results do not support our social integration and social influence hypotheses but provide evidence for selection: individuals with higher emotional well-being tend to have more strong-tied friends, and there are homophily processes regarding emotional well-being in strong-tied networks. Our study highlights the two-directional relationship between social ties and well-being, and demonstrates the importance of considering different tie strengths for various social processes.}


\begin{document}

\thispagestyle{empty}
\maketitle

\section{Introduction}

Social networks affect subjective well-being through various social mechanisms \citep{Berkman2000}. Due to the vast prevalence and indirect economic and social costs of low well-being, understanding these mechanisms is crucial for individuals, communities, and societies. However, research in the past decades has neglected to thoroughly examine social mechanisms on the micro level, in order to improve our understanding of \textit{how} social ties relate to individuals' well-being \citep{Thoits2011}.
\red{In this study, we investigate two groups of social mechanisms that explain the effect of social networks on subjective well-being}: \textit{social integration} and \textit{social influence}. \textit{Social integration} is conceptualized as the extent to which individuals participate in a broad range of social relationships and are embedded in dense networks \citep{Brissette2000}. \textit{Social influence} is defined as a ``change in an individual's thoughts, feelings, attitudes, or behaviors that results from interacting with another individual or a group'' \citep[][p. 4426]{Rashotte2006}. 
At the same time, social relationships do not develop independently of individuals' subjective well-being \citep{Schaefer2011}. Therefore, we investigate a third group of mechanisms focusing on \textit{social selection} processes, which capture that subjective well-being will affect the formation of social networks.

We propose that these three groups of processes -- social integration, social influence, and social selection -- operate differently depending on the strength of tie under investigation. 
On the one hand, strong-tied friends are emotionally closer to each other and are more likely to spend time together than individuals in weakly tied dyads; hence, we propose that social influence on subjective well-being happens mostly through strong ties. %
On the other hand, both strong and weak ties are important for social integration: the awareness of having many relationships, either weak or strong, induces a feeling of mattering to others, thereby nourishing the individual's sense of self-worth and acceptance within a community \citep{Thoits2011}. Thus, \textit{both} weak and strong ties should contribute to subjective well-being by providing a sense of belonging.
Related to social selection, we have two expectations. First, friendships should become stronger between individuals with similar levels of subjective well-being due to homophilic tendencies (i.e., bonding with similar others). \red{Second, individuals with higher subjective well-being should increase (and those with lower well-being should decrease) the number of one's strong and weak ties.} 
This argument is in line with Schaefer \textit{et al.} (\citeyear{Schaefer2011}) who argue that individuals with low well-being withdraw from their social networks.%

\red{In previous studies, weak and strong friendship ties have been operationalized as non-reciprocated and reciprocated ties, respectively \citep[e.g.,][]{Fujimoto2012, Carley1996}. 
\red{However, a distinction based on reciprocation is not necessarily a good proxy for tie strength. As an example, non-reciprocity might be the result of two individuals having a different understanding of what friendship means. 
At the same time, weak friendship ties can be  reciprocal if the individuals of a weak-tied dyad have a similar understanding of friendship.}
Other studies \citep[e.g.,][]{Hussong2002,Alexander2001} have distinguished between one best friend and all others. %
However, allowing only one best friend ignores the fact that there can be large individual differences in the number of strong ties (for instance, based on selection effects of well-being).}
We propose using a more fine-grained distinction between strong and weak ties. Our approach is based on the self-rated strength of relationships, where individuals are asked with whom they are acquainted  and to whom they relate as close friends. Referring to Granovetter's (\citeyear{Granovetter1973}) tie-strength definition, this approach allows to distinguish between weak and strong ties. A close friendship tie is more likely to entail a strong tie's properties (i.e., emotional intensity, intimacy, amount of time spent together, and reciprocal services) whereas acquaintance ties do not provide most of these properties and are therefore weak ties \citep{Granovetter1973}.  In that sense, Granovetter (\citeyear{Granovetter1983}) also uses the terms ``acquaintances'' and ``weak ties'' interchangeably. 

\red{To investigate hypotheses on these two sets of networks (weak and strong ties) simultaneously, we will use \textit{ordered} Stochastic Actor-Oriented Models \citep[ordered SAOMs; ][]{Snijders2010}\nocite{Snijders2010}. Since such ordered models have rarely been applied before, we will thoroughly discuss this method and demonstrate how it can be used to model ordered networks jointly.}

For examining the three aforementioned groups of mechanisms \red{on weak and strong ties}, we focus on an affect-based dimension of subjective well-being called \textit{emotional well-being}\footnote{
\linespread{1}The other dimension of subjective well-being entails memory-based evaluations of life such as life satisfaction \citep{Diener2008, Keyes2002}.}. Emotional well-being refers to the ``quality of an individual's everyday experience'' such as the experience of joy, happiness, positive affect, and the absence of sadness and negative affect  \citep[][p. 16489]{Kahneman2010a}.  It has been suggested that the experience of emotional well-being is only marginally related to socioeconomic factors such as wealth but rather is a socially facilitated phenomenon  \citep{Diener2010, Kahneman2010a}. \red{Thus, investigating the effects of social networks is crucial for understanding emotional well-being of individuals.}

This study seeks a better understanding of the social mechanisms behind emotional well-being by addressing three research questions. First, do high levels of social integration in weak- and strong-tied friendship networks lead to higher levels of emotional well-being? Second, is an individual's level of emotional well-being influenced by the well-being of those that he or she is strongly tied to? Third, how does an individual's emotional well-being affect his or her friendship selection patterns?%

To investigate the dynamic interplay between social ties and emotional well-being, we apply a longitudinal complete network design, which allows us to consider social integration, social influence, and social selection simultaneously.  
This is an important contribution of this study, since various scholars have called for complete network research on social integration \citep{Brissette2000, Kawachi2001}. This is because compared to egocentric network studies, analyses of complete networks can account for influences of the network structure and the relational perceptions and attributes of others.

 We use longitudinal friendship network data from the \textit{Friends and Family} study \citep{Aharony2011,Stadtfeld2015}. 126 individuals from a university-related graduate housing community in the US were part of this research project.  %
This empirical setting is unique in many ways. First, different qualities of friendship relations were measured in a complete network setting at various timepoints. These data allow us to identify a reasonable subset of weak and strong friendship ties and model their evolution over time. Second, affect was measured on a daily basis, which can be aggregated over time to a measure of emotional well-being \citep{Diener2008}. Compared to memory-based dimensions of subjective well-being (i.e., life satisfaction), this measure of affect is less prone to memory-biases such as duration neglect \citep{Kahneman1999} and recency effects \citep{Schwarz1999}. Third, as the participants moved into the community as couples, we can model changes in friendship networks and emotional well-being taking into account the role of the partner's emotional well-being. \red{From research on couples, we know that partners influence each other in their well-being \citep{Walker2011a}. Considering and controlling for such} partner effects contribute to the preciseness of the analysis and findings. Another interesting aspect of the empirical setting is that most individuals (91.4\%) moved to this community from other places within two years preceding the study. It is thus likely that this 
community served as an important part of their everyday social life in a phase of settling in. This makes social integration quite important for an individual's well-being, even though most friendship relations are relatively new.%

\red{As highlighted previously, a unique feature of this article is the application of }\textit{ordered} Stochastic Actor-Oriented Models \citep[ordered SAOMs; ][]{Snijders2010, Ripley}\nocite{Snijders2010, Ripley}. Ordered SAOMs simultaneously model the evolution of two (or more) networks in which each tie in one network (e.g., strong-tied network) is dependent upon the existence of the tie in another network (e.g., weak-tied network). Ordered SAOMs have rarely been applied in empirical studies so far, though they have various advantages when analyzing weighted ties, or subsets of ties. For example, hypotheses regarding mechanisms for strength-specific relationship dynamics can be tested simultaneously and cross-network effects (combining ties of different strength) can comprise a deeper understanding of the social selection mechanisms at play. Although the distinction between weak and strong ties has received much attention in various scientific disciplines \citep[e.g.,][]{Uzzi1999, Granovetter1973}\nocite{Uzzi1999, Granovetter1973}, the corresponding analytical methods are less advanced.
Some studies, for example, have analyzed strong and weak ties separately \citep[e.g.,][]{Baerveldt2004}. Ordered SAOMs allow for the simultaneous modeling of multiple ordered types of ties. Besides its theoretical contribution this article seeks to discuss the class of ordered SAOMs and show how it can be used to analyze hypotheses on different levels of ties.

In summary, this work contributes to existing research on social determinants and consequences of individuals' well-being in three major ways. First, it forms and tests theoretically based hypotheses on different levels of friendship ties.
Second, it disentangles social integration, social influence, and social selection mechanisms and thus provides more reliable results on the relationship between social ties and well-being. %
Third, by modeling weak and strong friendship ties jointly, it offers one of the first empirical examples utilizing the method of ordered SAOMs.

In the following sections, we outline the theoretical foundations of our hypotheses (Section 2) and provide more details on the empirical setting of this study (Section 3). Afterwards, we describe how ordered SAOMs can be applied to model the co-evolution of weak and strong friendship ties and emotional well-being (Section 4). Then, we present results of the ordered SAOMs (Section 5) and discuss our findings (Section 6). %

\section{Theory and hypotheses}

To be able to form theoretically based hypotheses, we rely on existing research on social integration, social influence, and social selection. We briefly summarize the most relevant findings on general subjective well-being as well as on the related construct of depression. We argue that the underlying social mechanisms for the maintenance of subjective and emotional well-being are similar to those for the absence of a depressive episode for three reasons. First, the main symptom for depression is low affect, which is also the core of low emotional well-being \red{\citep{Diener2010, Watson1988a}}. Similarly, the presence of positive subjective well-being includes the absence of mental disorders such as major depression \citep{WHO2003}.  Finally, depression and low well-being may be similar in that both are likely to be associated with having fewer and weaker social relationships, due to difficulties with social interaction (such as avoiding eye contact, impaired attention; Diener \textit{et al.}, 1999; Fredrickson, 2004; Segrin, 2000)\nocite{Diener1999, Fredrickson2004, Segrin2000}.%
\subsection{Research on social integration and well-being}

Social integration is a quality of the structural properties of one's social network \citep[e.g., reciprocity, density; ][]{Brissette2000}\nocite{Brissette2000}\footnote{
At the same time, social integration can be understood as the diversity of social roles that individuals take up (e.g., a parent, a co-worker). The quantity of such social roles has been consistently linked to well-being \citep[for a review see][]{Brissette2000}.}. 
In the following, we refer to these focal individuals as ``ego'' and to individuals that they are tied to as ``alters''.
It is argued that socially integrated individuals conform with normative expectations of the community \citep{Granovetter1973}. By meeting these expectations, individuals gain a sense of belonging, security, self-worth, and a positive self-identity and thus well-being \citep{Thoits2011,Cohen2004a}. Furthermore, socially integrated individuals have more access to information and social support \citep{Uzzi1999}.  In this article, we focus on two structural properties of social integration: \red{\textit{network size} and \textit{social cohesion}}\footnote{
\linespread{1}\red{Other scholars refer to what we call social cohesion as social embeddedness \citep{Ennett2006}, network cohesion \citep{Falci2009}, or egocentric density \citep{Ueno2005}.}}. 

\red{\textit{Network size} refers to the number of an individual's dyadic relationships  \citep{Brissette2000}}. %
  The perceived size of \red{individuals' own personal network} and the frequency of their interactions with alters has been found to be negatively associated with depressive symptoms \citep{Schaefer1981a}. \red{In an adolescent population, Ueno (\citeyear{Ueno2005}) has reported that the relationship between an individual's network size and his or her depressive symptoms is also negative and curvilinear.}

\red{\textit{Social cohesion}} captures network structures beyond dyads, including the extent to which an individual's alters are tied to each other  \citep{Friedkin2004}. Kadushin (\citeyear{Kadushin1982}) argues that connectedness among someone's alters can be beneficial for his or her well-being. Two mechanisms may play a role in this association. 
First, connected alters are more likely to share information about ego's well-being \citep{Kadushin1982}. Therefore, we argue that individuals in such dense structures are more effective and efficient in providing social support, which has been consistently linked to well-being \citep{Cohen2004a}. Second, cohesive triadic structures increase the level of trust in the group, which is also associated with higher subjective well-being of individuals \citep{Coleman1988, Helliwell2011}. 

To date, empirical studies on the role of social cohesion and well-being are relatively rare. As an existing example, Burt (\citeyear{Burt1987}) finds that the number of perceived positive relationships between someone's friends is positively associated with well-being. \red{Ueno (\citeyear{Ueno2005}) reports that the number of adolescent's school friends is negatively associated with depressive symptoms but finds no evidence for the role of social cohesion. }

Being socially integrated offers individuals a sense of belonging and access to various forms of social support, which in turn is beneficial for their well-being \citep{Brissette2000, Thoits2011}. 
In terms of tie strength, both strong and weak ties were argued to contribute to social integration. For instance, Lin and colleagues (\citeyear{Lin1999}) have shown that the network size within a community (i.e., a combination of weak and strong ties) is associated with higher perceived and received social support. In a similar vein, Granovetter (\citeyear{Granovetter1973}) has discussed that informational social support (i.e., finding a job) happens mostly through weak ties. Such informational types of social support can contribute to an individual's sense of belonging, which has been linked to fewer depressive symptoms \citep{Ueno2005}. Furthermore, the activation of social support should be more effective and efficient in socially cohesive structures of strong \textit{and} weak ties.
We therefore formulate the following hypotheses on the effect of perceived social integration on emotional well-being:  
\\ 
\\
\textit{Hypothesis 1a.} Individuals with more social ties in weak and strong friendship networks report higher levels of emotional well-being. 
\\
\\
\textit{Hypothesis 1b.} Individuals with higher levels of social cohesion (i.e., those maintaining more social ties towards others who are also tied together) in weak and strong friendship networks report higher levels of emotional well-being.

\subsection{Research on social influence and well-being}

Through mechanisms of social influence, the level of the adjunct alters' well-being can affect the individual's well-being.  %
Related studies mostly focus on the socialization of depressive symptoms in adolescents. %
Some of these find socialization effects \citep{VanZalk2010a, VanWorkum2013, Kiuru2012}, whereas others report no such effect \citep{Giletta2012, Pachucki2014}. These mixed findings suggest that potentially confounding mechanisms (e.g., social integration) or moderating effects (e.g., gender; van Zalk \textit{et al.}, 2010b\nocite{VanZalk2010b}) should be taken into consideration. %

Even though most previous studies are concerned with adolescents, social influence should play an important role in adults' emotional well-being as well. %
\red{Research indicates that adults with higher emotional well-being are more likely to provide others with social support, which has various positive effects on health and well-being for the recipient \citep{Cohen2004a, George1991}. }

Another mechanism of social influence is stated in Coyne's (\citeyear{Coyne1976a, Coyne1976}) \red{interactional theory of depression, which proposes that individuals with depressive symptoms induce negative affect in their interaction partners due to enhanced demands for emotionally comforting responses from them}. Research testing Coyne's theory has so far reported mixed evidence for the influence of affect \citep[for a meta-analysis see][]{Joiner1999}\nocite{Joiner1999}.
\red{A potential shortcoming of these studies might be that they are mostly lab-based and therefore only capture the effect of short-term interactions with single interaction partners. It is possible that repeated interactions with multiple others in a different affective state are necessary for social influence effects to be powerful.} %
In non-experimental settings, considering the \textit{type} of relationship between the individuals can be a key factor to explain this. Two individuals linked with strong ties are more likely to spend time together, discuss more emotionally relevant subjects, and share the intimacy of reciprocal disclosure \red{\citep{Granovetter1983}}. \red{Social influence mechanisms such as co-rumination \citep{Rose2002} and contagion of affect \citep{Coyne1976a} rely on the fact that these individuals interact with each other and are emotionally and mutually disclosing}. All these aspects of strong relationships provide more opportunities to be influenced by the emotional well-being of strong-tied alters \citep{Brechwald2011, Latan?1981}. %
Therefore, we derive the following hypothesis on the relationship between the emotional well-being of strongly tied individuals: 
\\ 
\\
\textit{Hypothesis 2.} Individuals' emotional well-being becomes more similar to the emotional well-being of their strong-tied alters.

\subsection{Social selection as an alternative mechanism}

So far, we have focused on the effect of social ties on emotional well-being. We have proposed two groups of mechanisms suggesting that that social ties in general positively affect emotional well-being (i.e., social integration), and individuals become similar to their friends over time (i.e., social influence). However, not only social ties affect well-being, but well-being also influences the quality and quantity of social ties one forms and maintains. The group of mechanisms is concerned with this latter direction is called social selection. Since processes of social integration/influence and social selection can result in the same outcome (higher well-being of those with more friends; similarity of friends), it is crucial to simultaneously focus on both kinds of processes. 
For social integration, an alternative hypothesis explains that individuals with higher well-being are more socially active and thus have a larger number of social ties.
For social influence, an alternative hypothesis explaining the similarity of friends is homophilic selection \citep[preference for similar others;][]{Mcpherson2001}\nocite{Mcpherson2001}. %

We assume that low emotional well-being is associated with reporting fewer weak and strong friendship ties because of its relation to decreased psycho-social functioning -- irrespective of whether the interaction partner is a friend, acquaintance, or a stranger \citep{Segrin2000,Fredrickson2004}. Individuals with low emotional well-being might therefore interact less with others and withdraw from their social network \citep{Schaefer2011}. %
Moreover, individuals with low emotional well-being might perceive social interactions as more negative due to unfavorable self-perceptions in social interactions associated with low well-being \citep{Gadassi2015, Gotlib1983}.
Hence, we would expect to see that individuals with low well-being withdraw from their friendship network (i.e., nominate fewer friends over time).

Another important social selection mechanism is homophily, which implies that individuals who are similar along dimensions relevant in a given context (e.g., gender, religion, depression, well-being) are more likely to be friends. A number of mechanisms explain the emergence of homophily in friendship networks. Those mechanisms can be roughly categorized into three classes. \textit{Baseline homphily} indicates an over-representation of homophilic ties due to overall composition, geographic segregation, social and organizational foci, or family ties \citep{Mcpherson2001}. \textit{Choice homophily} expresses homophilic preferences in individuals' friendship choices, for example, due to shared experiences and shared feelings of understanding \citep{McPherson1987}. \textit{Amplified homophily} is explained by the interplay of homophily with other social mechanisms such as transitivity that induce an over-representation of homophilic ties that is not captured by base-line homophily and choice homophily \citep{Stadtfeld2015}. By controlling for baseline homophily and potentially amplifying mechanisms explicitly in the empirical setting, we are able to interpret homophilic tendencies mostly in terms of choice homophily.

On the basis of choice homophily, one can argue that individuals with low well-being might seek for friends who are in a similar emotional state.
Alternatively, homophily on depressive symptoms (i.e., low emotional well-being) has been argued to arise because depressed individuals withdraw from their friendship network and find friends in similar peripheral network positions \citep{Schaefer2011}. 

Moreover, we expect emotional well-being homophily to be weaker for weak ties than for strong ties. Disclosure of low emotional well-being (e.g., talking about personal problems) is necessary for the homophilic process which is thus more likely to occur in strong-tied dyads  than weak-ties dyads. 
However, if low emotional well-being is disclosed to a weak-tied friend who is in a similar state and might provide more understanding due to similar experiences, this tie is more likely to get stronger. Such processes might result in emotional well-being homophily of strong ties but not of weak ties.

Given our assumptions about the relation between emotional well-being and social tie formation, we hypothesize \red{that}:
\\
\textit{Hypothesis 3a.} Individuals select others as strong-tied friends that report similar levels of emotional well-being.
\\ 
\textit{Hypothesis 3b.} Individuals with low emotional well-being report fewer weak social ties than individuals higher in emotional well-being.
\\
\textit{Hypothesis 3c.} Individuals with low emotional well-being report fewer strong-tied friendship ties than individuals higher in emotional well-being.
\\

\section{Empirical Setting}

\subsection{Participants}

The stated hypotheses are investigated using data from the \textit{Friends and Family }study \citep{Aharony2011, Stadtfeld2015}. %
\red{Of the initial 126 individuals in the study, we excluded nine individuals because they did not participate in the friendship questionnaire. The sample thus consisted of 117 individuals}. At the time of the study, all participants were in a long-term partnership and lived in the same US graduate housing community. In the sample, there were 55 partnership ties and seven individuals without their partner (six male and one female). Of the partnership ties, 53 were heterosexual ties and two male homosexual ties. 63 (53.8\%) participants were male. %
 Sixty-two individuals had children at the time of the study. The mean age of the sample was \red{28.5 years (\textit{SD} = 3.8 years, range 22 to 42 years)}, without two outliers of age 54 and 60 \red{and five missing values}. The five most frequent religions were 'atheist or no religion' (29.1\%),  Christian (16.2\%), Mormon (15.4\%), Catholic (12.8\%), and Jewish (12.0\%). \red{Three individuals did not provide information on their religion.} The major ethnicities were Asian (39.3 \%), White (38.5 \%), Hispanic (9.4 \%), Middle Eastern (6.0 \%), and Black (1.7 \%). \red{Three individuals did not provide information on their ethnicity.} 
 The majority of participants \red{(107; 91.4\%)} moved into the community within the two years preceding the study. Hence, the community was still emerging at the time of data collection.%

\subsection{Measures}
\subsubsection{Friendship}
Data were collected between September 2010 and May 2011. At four time-points (early September 2010, December 2010, March 2011, May 2011), interpersonal friendship relations and their quality were assessed with an online questionnaire. \red{Using a roster method, p}articipants were asked to evaluate the quality of their relationship with every other participant on an eight-point scale ranging from 'I don't know this person' (0) to 'this person is family or as close to me as a family member' (7). \red{The precise wording of the item was: 'For each name listed below, please fill in a numerical answer in the first column, then put an ``X'' in all other columns that apply.' Each column besides the first one, constituted of one friendship quality level.} Table~\ref{friendship_levels} presents these levels of these friendship qualities. This friendship quality rating allows us to identify a  subset of weak and strong ties of the social networks. There was no limit to the number of friendship nominations each participant could make. The first wave of data collection is dropped from the analysis as no well-being data had been collected at that time-point. 
We also used a number of demographic variables which were collected together with the first friendship ratings.  %
More details on the data collection and content can be found in Aharony \textit{et al.} (\citeyear{Aharony2011}) and Stadtfeld and Pentland (\citeyear{Stadtfeld2015}). 

\begin{figure}
    \includegraphics[scale=0.70]{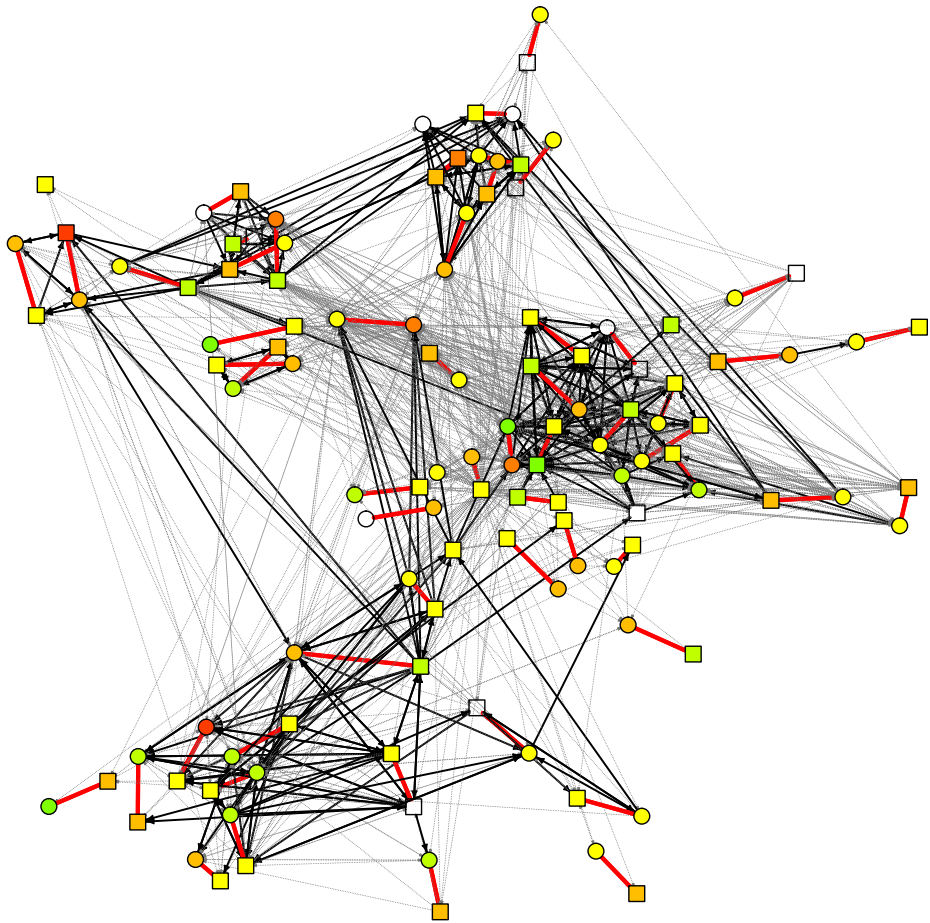}
    \caption[\red{Friendship network at T1 with weak, strong, and partnership ties}]
      {Friendship network at wave one. Partnership ties are marked red, weak ties with thin dotted lines, and strong ties as solid lines. Continuous node color represents well-being scores (red = low score, green = high score, white = missing). \red{Squared node shapes are male participants, whereas round node shapes are female.}}
    \label{netplot}
\end{figure}

\renewcommand{\baselinestretch}{1}
\begin{table}
  \caption{Levels of friendship ratings}
  \label{friendship_levels}
  \begin{minipage}{\textwidth}

    \begin{tabular}{L{1cm} L{8cm}}
      \hhline{==}
      Level & Friendship measure \\
           \hline
      0  &I don't know this person \\
      1 &I know of this person \\
      2 & This person is an acquaintance\\
      3 & This person is a friend (low ranking)	 \\
      4 &This person is a friend (medium ranking) \\
      5 & This person is a friend (high ranking)\\
      6 & This person is a close friend\\
      7 & This person is family or as close to me as a family member\\

     \hhline{==}
    \end{tabular}
  \end{minipage}
\end{table}
\renewcommand{\baselinestretch}{2}
We define weak ties as friendship nominations of strength \red{two (''This person is an acquaintance'') or larger which is in line with Granovetter's (1983) terminology of weak ties\nocite{Granovetter1983, Granovetter1973}}. Strong ties are defined as nominations of strength five (''This person is a friend (high ranking)'') or larger\footnote{ We tested the robustness of these two cutoff values by additionally conducting analyses with weak ties cut off at level three (``This person is a friend (low ranking'') and strong ties cut off as six (``This person is a close friend''). A summary table of all robustness checks can be found in the appendix Table ~\ref{robust}. }. \red{For the analysis of ordered SAOMs with RSiena \citep{Ripley}, we then created two dichotomized, asymmetric adjacency matrices, of which the strong ties network is a subset of the weak-tied network\footnote{
\linespread{1}The class of ordered SAOMs requires networks to be nested within each other.}.}  Table~\ref{net_desc} entails descriptive statistics of these weak and strong friendship networks. The density and average degree increase with time. Jaccard indices indicating the proportion of stable ties between two subsequent observations range between .61 and .73, which are considered good amounts of change for an analysis with SAOMs \citep{Ripley}. The number of individuals who did not fill out the friendship questionnaire increased from nine in wave one to 35 in wave three. Figure~\ref{netplot} shows a visualization of the weak and strong friendship networks, and the partnership networks at wave one. Weak ties in Figure~\ref{netplot} are represented as thin dotted lines, strong ties as black lines, and partnership ties as thick red lines. The color of each node corresponds to the individuals' continuum of emotional well-being, where red indicates low values, green indicates higher values, and white represents missing values. \red{A visualization of the networks at wave three can be found in the appendix (Figure~\ref{netplot2})}.

\renewcommand{\baselinestretch}{1}
\begin{table}
  \caption{Descriptive statistics of weak and strong friendship networks }
  \label{net_desc}
  \begin{minipage}{\textwidth}
\begin{tabular}{L{2cm}
L{3cm}
S[table-format=3.2,table-align-text-post = false]
S[table-format=2.3,table-align-text-post = false,add-integer-zero=false]
S[table-format=3.2,table-align-text-post = false]
S[table-format=2.3,table-align-text-post = false,add-integer-zero=false]
S[table-format=3.2,table-align-text-post = false]}
\hhline{=======}
 &  & T1 &  & T2 &  & T3 \\ \hline

Weak & Average degree & 13.9 &  & 15.0 &  & 16.9 \\
 & Density & 11.2\% &  & 12.0\% &  & 13.6\% \\
 & Jaccard index &  & .71 &  & .73 \\ \hline

 Strong& Average degree & 3.1 &  & 3.4 &  & 4.0 \\
 & Density & 2.5\% &  & 2.8\% &  & 3.2\% \\
 & Jaccard index &  & .61 &  &.68 \\ \hline
  & Ties missing & 17.3\% &  & 27.6\% &  & 25.2\% \\
\hhline{=======}
\end{tabular}

 \end{minipage}
\end{table}

\renewcommand{\baselinestretch}{2}

To examine whether missing friendship ratings occurred systematically, we conducted a series of logistic regression analyses for each of the weak and strong friendship networks. In every model, the dependent variable was whether the observation was missing. The explanatory variables consisted of a combination of characteristics of an individual's network position in the previous network wave and and measures of emotional well-being in the previous period (i.e. mean score its and variance over time). None of these variables consistently explained missing friendship ratings. %

\subsubsection{Emotional well-being}
On a daily basis, participants were asked to rate their momentary affect on a seven-point scale with the following item: ``how happy were you on \textit{X}?'', where \textit{X} represents the current weekday (e.g., Friday).  The scale ranged from 'very unhappy' (1) to 'very happy' (7). This was done using a mobile phone application. We compute the mean of these happiness scores over a period of 30 days before each network assessment to calculate an aggregate measure for emotional well-being\footnote{
\linespread{1} \red{We tested the robustness of this 30 day threshold by additionally conducting all analyses with alternative cutoffs of 20 and 40 days. A summary table of all robustness checks can be found in the appendix Table ~\ref{robust}.}}. %
Emotional well-being is assessed as the mean of individuals' momentary or daily affect over several weeks as suggested by Diener and Lucas (\citeyear{Diener2008}). A total of 4'056 well-being ratings are aggregated into 274 individual observations over time. Ideally, positive and negative dimensions of emotional well-being are measured with separate scales as these dimensions are argued to be partially independent \citep{ Watson1988,	Huppert2010}. However, in a more simplified way, positive and negative affect can be reasonably treated as the two ends of one dimension, one end of the scale representing a negative emotion ('very unhappy') and the other end a positive emotion ('very happy')%
\footnote{
\linespread{1}\red{In the most prominent affect measures -- the Positive and Negative Affect Schedule \citep{Watson1988} -- the extent to which an emotion is experienced is measured on a scale ranging from 'very slightly or not at all' to ``extremely'. In that case, the low end does not represent the negative dimension, whereas in our case it does.}}.

There is no significant correlation between the number of ratings used for the aggregation and the aggregated score of emotional well-being (\textit{p} $>$ .10), or between the individuals' variance and the aggregated score of emotional well-being (\textit{p} $>$ .10). Furthermore, the mean variance of the individuals' ratings is low (\textit{M}$_{var T1}$ = 1.08, \textit{SD}$_{var T1}$ = 0.76; \textit{M}$_{var T2}$ = 0.94, \textit{SD}$_{var T2}$ = 0.84; \textit{M}$_{var T3}$ = 0.76, \textit{SD}$_{var T3}$ = 0.65). Given the temporal stability of these ratings, an aggregation seems valid. Table~\ref{happy_desc} shows the descriptive properties of these emotional well-being ratings.

\renewcommand{\baselinestretch}{1}
\begin{table}
  \caption{\red{Descriptive statistics of aggregated emotional well-being scores}}
  \label{happy_desc}
  \begin{minipage}{\textwidth}
\begin{tabular}{l C{1.5cm} C{1.5cm} C{1.5cm}C{1.5cm}C{2.5cm}C{1.5cm}C{1.5cm}} \hhline{=======}
&\textit{M}$_{N}$&\textit{SD}$_{N}$&\textit{M}$_{agg}$&\textit{SD}$_{agg}$& missings$_{agg}$ (\%)&\textit{M}$_{indVar}$\\ \hline
T1 & 18.33 & 4.71 & 4.92 & 0.86 &  9 (7\%) & 1.09 \\
T2 & 20.99 & 10.50 & 4.73 & 0.93 & 28 (22\%) & 0.94 \\
T3 & 21.96 & 10.16 & 4.85 & 0.91 & 46 (37\%) & 0.76 \\
\hhline{=======}
\end{tabular} 
 \end{minipage}
 \newline \\
\begin{tablenotes}
      \item \textit{Note.} \red{\textit{N(participants)} = 117, \textit{N(ratings}) = 4'056, \textit{M}$_{N}$ = Mean number of individual's ratings used for aggregation, \textit{M}$_{agg}$ = Mean value of the aggregated ratings, \textit{M}$_{indVar}$ = Mean variance of each individual's ratings used for aggregation.}
    \end{tablenotes}
\end{table}
\renewcommand{\baselinestretch}{2}

We further examine patterns of missing emotional well-being scores to determine whether these missing values are related to emotional well-being, and could thus bias our analysis. In a series of logistic regression analyses, values of previous emotional well-being scores and degree measures of neither weak- nor strong-tied networks predict future missing values of emotional well-being (\textit{p} $>$ .10). Hence, we assume in our analyses that emotional well-being scores were missing completely at random. 

\section{Methods}

The longitudinal data on social networks and well-being are analyzed using Stochastic Actor-Oriented Models for ordered networks \citep[ordered SAOMs; ][]{Snijders2010}.  This method allows investigating the multi-mechanistic and dynamic hypotheses proposed. SAOMs simultaneously model changes in network ties (i.e., strong and weak ties) and individual attributes (i.e., emotional well-being). SAOMs allow to jointly test the effects of social integration and social influence on individual well-being, while taking into account that emotional well-being may affect changes of network ties \citep{Steglich2010a}. 

	  \subsection{Stochastic Actor-Oriented Models} 
SAOMs are used to analyze longitudinal data of social networks and individual attributes measured at discrete points in time. In SAOMs, network and attribute changes are modeled jointly as separate dependent variables. Under which circumstances ties and individual attributes change is characterized by so-called effects that can, for example, capture the tendencies of individuals to form ties within triadic structures or with similar others. These effects are used as explanatory variables in the analysis. To model network changes, a variety of structural effects (e.g., the tendency to reciprocate ties), dyadic characteristics (e.g., the tendency to become friends with neighbors), and nodal attributes (e.g., gender homophily) can be included. To model individual attribute changes, numerous structural effects (e.g., an individual's reciprocated degree) and effects of alters' attributes (e.g., the tendency to assimilate to the alters' emotional well-being) are used. The effects relevant for our analyses are described in more detail in Section~\ref{model specification}. Ripley \textit{et al.}(\citeyear({Ripley}), section 5 provide a broad overview of potential model specifications.
Based on the included effects,  simulations of a continuous-time change process are carried out so that the characteristics of the simulated networks match those of the subsequent, empirical observation. In so-called \textit{mini steps}, actors assess all possible outcomes of potential changes of at most one tie variable and at most one attribute level to evaluate whether an outgoing tie should be created, dropped, or maintained  or whether the attribute should be increased, decreased, or kept at the previous level. %
The frequency in which actors are selected to execute the mini step is modeled through the \textit{rate function}. The evaluation of the network and attribute of the focal actor is modeled through the \textit{objective function}. The objective functions of network change and attribute change for actor \textit{i} are defined as linear functions:
	  \begin{equation}
	  f_i^{net} (\theta, x, z) = \sum_k \theta_k^{net} s_{ik}^{net} (x, z)
	  \end{equation}
	  	  \begin{equation}
	  f_i^{beh} (\theta, x, z) = \sum_k \theta_k^{beh} s_{ik}^{beh} (x, z).
	  \end{equation}
	  
\noindent Parameters $\theta_k^{net}$ and $\theta_k^{beh}$ correspond to effect functions $s_{ik}^{net} (x, z)$ and $s_{ik}^{beh} (x, z)$, which were included in the model as explanatory variables. $\theta_k^{net}$ and $\theta_k^{beh}$ are subject to statistical inference.
Variables $x$ and $z$ represent the network ($x$) and individual attributes ($z$) of all actors at the time of a mini step. Actors in a given mini step are most likely to change the outgoing ties or attributes in the way that leads to the highest values of $f_i^{net}$  and $f_i^{beh}$ respectively. The choice between different mini steps is modeled as a multinomial model. Snijders \textit{et al.} (\citeyear{Snijders2010}) explain the mathematical model in more detail.

We use the method of moments for parameter estimation, in which the parameters of the network and attribute statistics are modeled to fit the values of future and past network and attribute measurements. SAOMs are estimated with the RSiena package \citep{Ripley}. Further statistical specifications of SAOMs are described in Snijders (\citeyear{Snijders2001}), Snijders \textit{et al.} (\citeyear{Snijders2010}),  and Steglich \textit{et al. }(\citeyear{Steglich2010a}). 

\subsection{Modeling weak and strong ties with \textit{orderd }SAOMs}

SAOMs allow the simultaneous modeling of two (or more) ordered networks, where the evaluation of a tie in one network (i.e., strong friendship network) is dependent upon the presence of the tie between the same actors in another network (i.e., weak friendship network). 
\red{Therefore, the strong friendship network needs to be nested within the weak friendship network.} The network objective function (see Equation 1) is replaced by two objective functions -- one for each network. SAOMs for ordered networks include a dependency in the simulations between the weak- and strong-tied networks: Individuals can only form a strong tie with individuals to whom they are already tied to in the weak-tied network and they can only completely dissolve weak ties. In case a strong tie is created between two previously unconnected individuals, a weak tie has to be created first and strong ties can only be dissolved completely by transforming them to a weak tie first.

By applying ordered SAOMs, we can test social mechanisms (e.g., homophily, transitivity) separately on weak and strong networks.
Furthermore, ordered SAOMs allow including and testing cross-network effects. These capture the tendency to form or maintain ties in one network depending on ties in the other network. For example, we examine the effect of a mutual tie in the weak friendship network on the creation and maintenance of a strong tie. Such cross-network model specifications are similar to SAOMs for multiplex networks \citep[e.g.,][]{Pal2015,Ripley,Snijders2013}\nocite{Pal2015,Ripley,Snijders2013}.

Are these assumptions about ties moving through an ordered sequence from weak to strong and strong to weak reasonable?
In our data we find that most new ties in the strong friendship network were weak ties before ($N_{T1-T3}$ = 113) and were rarely non-existing before ($N_{T1-T3}$ = 39). Most ties that disappeared from the strong-tied network became weak ties ($N_{T1-T3}$= 81) compared to becoming non-existent ($N_{T1-T3}$ = 18). Ties that evolved from non-existing to strong also moved through this sequence in the simulation and most likely in reality (every close friend is first an acquaintance, even if only for a short period).

\subsection{Model specification}\label{model specification}

In the following paragraphs we describe the effects that we use to specify our models. First, we discuss effects that relate to change in the strong and weak tied networks, and second, we discuss effects that relate to change of emotional well-being. Some effects that we discuss operate across these two levels.

\subsubsection{Strong and weak network change model} 

\red{Individuals change their strong and weak network ties based on the prevalence of network structures and individual attributes. }
\red{Both sub-models (i.e., weak and strong friendship networks) were specified equally to ensure comparability between the sub-models.}
	  
The baseline model consists of effects testing the density, reciprocity, \red{indegree-popularity, outdegree-popularity, outdegree-activity, transitive triples,} and a \red{ transitive reciprocated triplets} effect. 
\red{The three degree related effects -- indegree-popularity, outdegree-popularity, and outdegree-activity -- were included to control for the tendency of individuals to be more popular or active based on their in- or outdegree.}
The \red{transitive triples} effect is included to account for triadic clustering. A transitive reciprocated triplets is added to capture the tendency not to reciprocate ties within \red{transitive triples} as an alternative to the commonly used three-cycle effect \citep{Block2015a}.  %

 As friendships form in dependence with partnership ties \citep{Stadtfeld2015}, we include a dyadic partnership-covariate to account for the following effects: (1) Individuals have the tendency to become friends with the friends of their partner; (2) individuals have the tendency to become friends with the partner of a friend; (3) individuals show the tendency to close a couple-four-cycle (two couple ties, one friendship tie) if the alter is the same gender but not of the different gender.
 These effects were introduced by Stadtfeld and Pentland (\citeyear{Stadtfeld2015}) where the mathematical details are further explicated. %
	 
We further include effects that capture the tendencies of individuals who are high in emotional well-being to nominate more friends (\textit{emotional well-being ego}) and being nominated more frequently (\textit{emotional well-being alter}). We further test the tendency to create or maintain homophilic friendship ties with individuals of similar emotional well-being (\textit{emotional well-being similarity}). We include a number of additional homophily effects. These are the tendencies to have friends who have the same gender, the same religion, the same ethnicity, the same parental status (having children or not), are similar in age, and who are neighbors (propinquity). 
These effects were also chosen in line with the model specification of Stadtfeld and Pentland (\citeyear{Stadtfeld2015}).

	   \subsubsection{Emotional well-being model} 
Individuals' levels of emotional well-being is affected by individual characteristics, their network position as well as the emotional well-being of those they are connected to.
Network size as a form of social integration is operationalized as the number of reciprocated friendship ties. We focus on the reciprocated degree because mutual friends should contribute most to someone's sense of belonging and can be expected to provide social support. Social cohesion is measured with the number of dense triads -- triads that include five or six ties -- that individuals are embedded in. This means that in these triads, everyone is tied to everyone else and at most one dyad can be non-reciprocal. Being part of many dense triads suggests that individuals have many friends who are also friends with each other. \red{It is important to note that this effect only captures local (i.e., triadic) social cohesion. Currently, there is no effect in the RSiena framework available that captures a more global measure of social cohesion.}
Social influence is operationalized as the tendency of individuals to become similar to the average level of emotional well-being of their friends. In the model, we include influence effects of weak and strong ties, and of an individual's romantic partner. As RSiena does not support continuous dependent variables, we round each aggregated well-being score to the closest \red{half} number and then recode them to integers (e.g., 2.30 $\rightarrow$ 2.5 $\rightarrow$ 4; 4.76 $\rightarrow$ 5.0 $\rightarrow$ 9) to transform the measure into an ordinal categorical variable. For the internal use in RSiena instead of a scale from 1 to 6 we now have a scale from 1 to 13, where previous half distances count as a whole distance between values. 
Individual covariates such as gender could further be included in a SAOM to model change in an individual variable. The shape of the distribution of emotional well-being among the participants is modeled by including a linear and a quadratic term.

\section{Results}

First, we estimate a baseline model with parameters explaining the change of strong and weak friendship ties and  parameters explaining the overall distribution of emotional well-being values. Then, we add effects in a stepwise procedure until in a final model all effects expressing our hypotheses are tested simultaneously. We use a stepwise approach to show how parameter sizes and significance levels (of those effects explaining tie-selection and changes in emotional well-being) change when adding social integration and social influence effects. Hence, in the first two steps, we add effects which operationalize the social integration hypotheses (step one: reciprocated degree, step two: dense triads). In the third step, we include the average similarity influence effect of strong ties. In the fourth step, we add the average similarity influence effects of partners \footnote{
\linespread{1}\red{As partners share experiences in many domains of life, we can assume that their emotions and behaviors are not independent from each other. In fact, there is a substantive body of research on how partners influence each other \citep[actor-partner interdependence models; ][]{Cook2005}\nocite{Cook2005}. For instance, Walker et al. (2011) \nocite{Walker2011a} found that husbands over time become more similar to their wife’s well-being. For this reason, we control for the partner's influence on emotional well-being in the model.}}. In the fifth and final step, we include the average similarity effect of weak ties. Hence, in total six models are estimated. The results of the baseline model, the estimates of the effects added at each step, and the final model are presented in Table~\ref{SAOMs}. The changes of previously included parameter estimates at each step are not presented but reported when relevant.
In general, positive $\theta$ parameters represent the tendency of individuals to report ties contributing to the given effect (e.g., to reciprocate a tie) or to report higher levels of emotional well-being in case of higher values of the effect (e.g., individuals become more similar to their friends).  %
The estimates of the baseline model correspond well with the results of Stadtfeld and Pentland (\citeyear{Stadtfeld2015}), who used the same data (but different friendship definitions) and a similar baseline model specification. 
\renewcommand{\baselinestretch}{1} %

\begin{table}
\begin{scriptsize}

  \caption{\red{Results of ordered SAOMs on weak and strong friendship networks, and emotional well-being}}
  \label{SAOMs}
  \hspace*{-1cm}
\begin{tabular}{ %
l
S[table-format=3.4,table-align-text-post = false]
S[table-format=3.4,table-align-text-post = false]
S[table-format=3.4,table-align-text-post = false]
S[table-format=3.4,table-align-text-post = false]
S[table-format=3.4,table-align-text-post = false]
S[table-format=3.4,table-align-text-post = false]
S[table-format=3.4,table-align-text-post = false]
S[table-format=3.4,table-align-text-post = false]
S[table-format=3.4,table-align-text-post = false]}
\hhline{==========}
 & \multicolumn{4}{c}{Stepwise selection model}   & \multicolumn{4}{c}{Final model} \\ \cmidrule{2-9}
 & \multicolumn{2}{c}{Weak}   & \multicolumn{2}{c}{Strong}   & \multicolumn{2}{c}{Weak}   & \multicolumn{2}{c}{Strong} \\ \cmidrule{2-9}
Baseline model & \multicolumn{1}{c}{$\theta$} & \multicolumn{1}{c}{\textit{S.E.}} & \multicolumn{1}{c}{$\theta$} & \multicolumn{1}{c}{\textit{S.E.}} & \multicolumn{1}{c}{$\theta$} & \multicolumn{1}{c}{\textit{S.E.}} & \multicolumn{1}{c}{$\theta$} & \multicolumn{1}{c}{\textit{S.E.}} \\
\hline 
rate (period 1) & 10.26*** & 0.78 & 3.93*** & 0.49 & 10.26*** & 0.75 & 3.93*** & 0.48 \\
rate (period 2) & 9.67*** & 0.65 & 3.25*** & 0.38 & 9.66*** & 0.65 & 3.25*** & 0.41 \\
outdegree (density) & -3.16*** & 0.13 & -2.82*** & 0.39 & -3.17*** & 0.14 & -2.83*** & 0.40 \\
reciprocity & 0.99*** & 0.16 & 3.46*** & 0.51 & 0.99*** & 0.17 & 3.44*** & 0.51 \\
transitive triples & 0.06*** & 0.01 & 0.39*** & 0.08 & 0.06*** & 0.01 & 0.38*** & 0.08 \\
transitive reciprocated triplets & 0.01 & 0.02 & -0.42** & 0.13 & 0.01 & 0.02 & -0.41*** & 0.12 \\
indegree - popularity & 0.03*** & 0.00 & 0.01 & 0.04 & 0.03*** & 0.00 & 0.01 & 0.04 \\
outdegree - popularity & -0.03*** & 0.01 & -0.14** & 0.04 & -0.03*** & 0.01 & -0.14** & 0.04 \\
outdegree - activity & 0.01*** & 0.00 & 0.02 & 0.01 & 0.01*** & 0.00 & 0.02 & 0.01 \\
propinquity & 0.67*** & 0.12 & 1.08*** & 0.24 & 0.67*** & 0.12 & 1.07*** & 0.22 \\
friends with the friends of their partner  & 1.28*** & 0.13 & -0.44 & 0.37 & 1.28*** & 0.13 & -0.41 & 0.36 \\
friends with the partner of a friend  & 0.80*** & 0.06 & 0.83*** & 0.17 & 0.80*** & 0.06 & 0.82*** & 0.15 \\
couple-four-cycle closure & -0.30 & 0.20 & 1.44*** & 0.37 & -0.30 & 0.19 & 1.42*** & 0.36 \\
couple-four-cycle closure with same gender & 0.13 & 0.21 & 0.20 & 0.50 & 0.13 & 0.21 & 0.21 & 0.50 \\
same gender & 0.30*** & 0.08 & 0.20 & 0.23 & 0.30*** & 0.08 & 0.20 & 0.22 \\
same ethnicity & 0.23*** & 0.07 & 0.28 & 0.20 & 0.24*** & 0.07 & 0.27 & 0.19 \\
age similarity & -0.45* & 0.20 & 0.58 & 0.88 & -0.45* & 0.21 & 0.58 & 0.88 \\
same parental status & 0.07 & 0.06 & 0.15 & 0.17 & 0.07 & 0.06 & 0.16 & 0.17 \\
same religion & 0.21* & 0.09 & -0.06 & 0.20 & 0.21* & 0.09 & -0.05 & 0.20 \\
emotional well-being alter & 0.04 $\dagger$ & 0.02 & -0.14 $\dagger$ & 0.08 & 0.04$\dagger$ & 0.02 & -0.16* & 0.08 \\
emotional well-being ego (\textit{H3b, H3c})& -0.04 & 0.03 & 0.24* & 0.09 & -0.04 & 0.03 & 0.24** & 0.09 \\
emotional well-being similarity (\textit{H3a})& -0.18 & 0.40 & 3.05* & 1.28 & -0.17 & 0.39 & 3.14* & 1.35 \\
mutuality with weak &  &  & -0.27 & 0.39 &  &  & -0.30 & 0.42 \\
\hline \\
& \multicolumn{2}{c}{Emotional Well-Being}   &  && \multicolumn{2}{c}{Emotional Well-Being} \\
& \multicolumn{1}{c}{$\theta$} & \multicolumn{1}{c}{(SE)}  &&& \multicolumn{1}{c}{$\theta$} & \multicolumn{1}{c}{(SE)} \\
\hline \\	
 Baseline Model \\
rate (period 1) & 5.53*** & 0.97 &  &  & 5.65*** & 1.34 \\
rate (period 2) & 2.77*** & 0.58 &  &  & 2.84*** & 0.64 \\
linear shape & -0.04 & 0.05 &  &  & 0.00 & 0.11 \\
quadratic shape & -0.06*** & 0.02 &  &  & 0.01 & 0.06 \\
Step 1 \\
reciprocated degree (weak; \textit{H1a}) & 0.00 & 0.01 &  &  & -0.01 & 0.03 \\
Step 2 \\
dense triads (weak; \textit{H1b}) & -0.02 & 0.03 &  &  & 0.00 & 0.00 \\
Step 3 \\
average similarity (strong; \textit{H2}) & 2.37$\dagger$ & 1.28 &  &  & 0.69 & 3.99 \\
Step 4 \\
average similarity (couple) & -0.23 & 1.77 &  &  & -0.05 & 1.79 \\
Step 5 \\
average similarity (weak) &  &  &  &  & 4.16 &	5.91 \\

\hhline{==========}
\end{tabular}
	\hspace*{-1cm}
	\newline
	\end{scriptsize}
\begin{tablenotes}	
      \item \textit{Note.} \textit{N} = 117. The ordered models entail networks $X_{weak}$ and $X_{strong}$ where  $X_{strong}$ $\subset$ $X_{weak}$. All models were estimated with the methods of moments an 4'000 iterations in phase three. The models converged according to the t-ratios for convergence and the overall maximum
convergence ratio criteria suggested by Ripley \textit{et al.} (\citeyear{Ripley}). \red{Goodness of fit was adequate for all models.}
      \\ $\dagger$ \textit{p} $<$ .10, * \textit{p} $<$ .05, ** \textit{p} $<$ .01, *** \textit{p} $<$ .001. Two-sided \textit{p}-values.  
    \end{tablenotes}

\end{table}

\renewcommand{\baselinestretch}{2}

\subsection{Social integration processes}
First, we focus on individuals' network size. Hypothesis 1a predicts that maintaining more social relationships leads to higher emotional well-being. Hence, we include the \textit{reciprocated degree} effect in the model explaining emotional well-being. As this parameter at step 1 in Table~\ref{SAOMs} reveals, there is no evidence for this association in our weak-tied network ($\theta$ = 0.00, \textit{p} $>$ .10). Therefore, we do not find support for hypothesis 1a.

However, there are other effects that could potentially capture network size: in- and outdegrees, or having no in- or out-ties at all (in- /outdegree isolate). In a further explorative analysis, we therefore test whether any of these effects are significant predictors of emotional well-being\footnote{
\linespread{1}As the outdegree isolate effect was not present in the current RSiena version \citep{Ripley}, we include the outdegree isolate effect in the following way:
\begin{equation}
s^{\rm beh}_{i}(x, z) = z_i I\{x_{i+} = 0 \}
\end{equation}

 where $s^{\rm beh}_{i}(x, z)$ is the behavior statistics of actor \textit{x} for behavior \textit{z} and \textit{i+} is the tie from actor \textit{i} to all alters. $I\{A\}$ denotes a dummy variable for condition $A$.}.  
 For this, we apply score-type tests following the methods of Schweinberger (\citeyear{Schweinberger2012}), which test whether estimates of a non-modeled effect differ from zero. The score-type tests reveals that none of these effects is significantly associated with emotional well-being \red{(\textit{p} $>$ .10)}. 

Hypothesis 1b states that \red{individuals in socially cohesive structure (i.e., being part of many dense triads) report higher levels of emotional well-being.} Therefore, in the second step, we look at the effect of dense triads in weak-tied networks on emotional well-being. We find no association for this dense triads effect in our data (step 2 in Table~\ref{SAOMs}, $\theta$ = -0.02, \textit{p} $>$ .10). Therefore, we do not find support for hypothesis 1b in our analysis. 

\subsection{Social influence processes}

Step 3 of Table~\ref{SAOMs} focuses on our social influence hypothesis (\textit{H2}) predicting that individuals assimilate to their strong-tied alters' emotional well-being. Here, we find a borderline significant \textit{average similarity} parameter in the strong-tied network ($\theta$ = 2.37, \textit{p} $<$ .10). This might suggest that indeed strong-tied alters influence each other's emotional well-being. \red{Therefore, we find weak support for hypothesis two.} Compared to step 2, the estimate for the quadratic shape of emotional well-being is now non-significant ($\theta$ = -0.03, \textit{p} $>$ .10). This change might suggest that strong-tied alters' influence accounts for individual's tendencies towards less extreme values of emotional well-being. 

For the next step, we take into account that individuals have the tendency to report similar values of emotional well-being as their romantic partner (step 4 in  Table~\ref{SAOMs}). In our model the \textit{average similarity (couples)} effect is not significantly associated with emotional well-being ($\theta$ = -0.23, \textit{p} $>$ .10). At this step, the average similarity parameter of strong ties is non-significant ($\theta$ = 2.70, \textit{p} $>$ .10). %
To strengthen our theoretical assumptions about influence effects of weak and strong ties, we take into account that social influence of emotional well-being could happen on weak ties as well. Thus, for the final model we add the \textit{average similarity (weak)} effect. As the average similarity effect of strong ties is already in the model, this effects captures the social influence of weak ties only, even though the ordered SAOMs assume that strong ties are a subset of the weak-tied network. As expected, the weak ties average similarity parameter is not significantly associated with individuals' emotional well-being ($\theta$ = 4.16, \textit{p} $>$ .10). Also, the average similarity effect of strong ties remains non-significant when including the average similarity parameter of weak ties ($\theta$ = 0.69, \textit{p} $>$ .10). \red{The disappearance of the influence effect of strong ties might be due to insufficient statistical power and a strong collinearity with the influence effect of partners and weak ties. }
\red{With a multi-parameter Wald test (for details see Ripley \textit{et al.}, 2015) we also test whether all three influence effects together significantly contributed to the emotional well-being evolution. We report no joint significant influence of these three effects ($\chi^2$(3) = 3.96, \textit{p} $>$ .10).}

\subsection{Social selection processes}

We further assume that weak-tied individuals are more likely to become and remain strong-tied friends if they report similar emotional well-being (homophily hypothesis 3a) and that individuals with lower emotional well-being nominate fewer individuals in weak (hypothesis 3b) and strong friendship networks (hypothesis 3c). Estimates in Table~\ref{SAOMs} show that there is a significant effect for individuals to nominate others with similar emotional well-being in the strong-tied network ($\theta$ = 3.14, \textit{p} $<$ .05) but not in the weak-tied network ($\theta$ =  -0.17, \textit{p} $>$ .10). Again, this suggests that similarities are important when choosing and keeping close friends, but they are not as relevant for weaker relationships. We thus report support for hypothesis 3a.

We find a significant tendency of individuals with higher emotional well-being to nominate more strong-tied friends than individuals with lower emotional well-being ($\theta$ = 0.24, \textit{p} $<$ .01). This is not the case for weak ties ($\theta$ = -0.04, \textit{p} $>$ .10). Hence, we find evidence for hypothesis 3c, but our results do not support hypothesis 3b.

\red{Furthermore, we observe a borderline significant tendency of individuals to create and maintain weak ties with alters that report high emotional well-being ($\theta$ = 0.04, \textit{p} $<$ .10). Interestingly, the opposite effect seems to be the case in the strong-tied network, where individuals tend to avoid nominating alters with high levels of emotional well-being ($\theta$ = -0.16, \textit{p} $<$ .05).}

\red{To examine the three selection effects driving the evolution of the strong tied network in more detail, we compute the relative gain of the objective function on each level of ego's emotional well-being to nominate alters on each levels of emotional well-being. Ego's gain of the objective function is computed in the following way\footnote{
\linespread{1}More information on how to construct and interpret an ego - alter selection table can be found in the RSiena Manual \citep{Ripley}.}:
\begin{equation}f_{i,j}^{net} (\theta, z) = \theta_{\rm ego}\, (z_i - \bar z)
              \, + \, \theta_{\rm alter}\,  (z_j - \bar z) \, + \,
        \theta_{\rm sim} \,  \Big( 1 - \frac{\vert z_i - z_j \vert}{\Delta_Z}
                     - \widehat{{\rm sim}^z} \Big) \end{equation}
where  $\theta_{\rm ego}$,$\theta_{\rm alter}$, and $\theta_{\rm sim}$ denote the parameter estimates for the ego, alter, and similarity selection effects.  $\Delta_Z$ denotes the observed range of the emotional well-being variable $z$ and  $\widehat{{\rm sim}^z}$ is the mean of all similarity scores.
  For instance, the gain of an individual with emotional well-being of 2.5 (RSiena categorical value = 4) to nominate an individual with values of 6.5 (RSiena categorical value = 12) is $ 0.24(4-8.59) - 0.16(12-8.59) + 3.14\Big( 1 - \frac{\vert 4 - 12 \vert}{11} - 0.79 \Big) = -3.28$. Controlling for all other effects in the model, individuals with values of 2.5 are unlikely to nominate alters with an emotional well-being value of 6.5. }
  
  \renewcommand{\baselinestretch}{1}
\begin{figure}
    \includegraphics[scale=0.55]{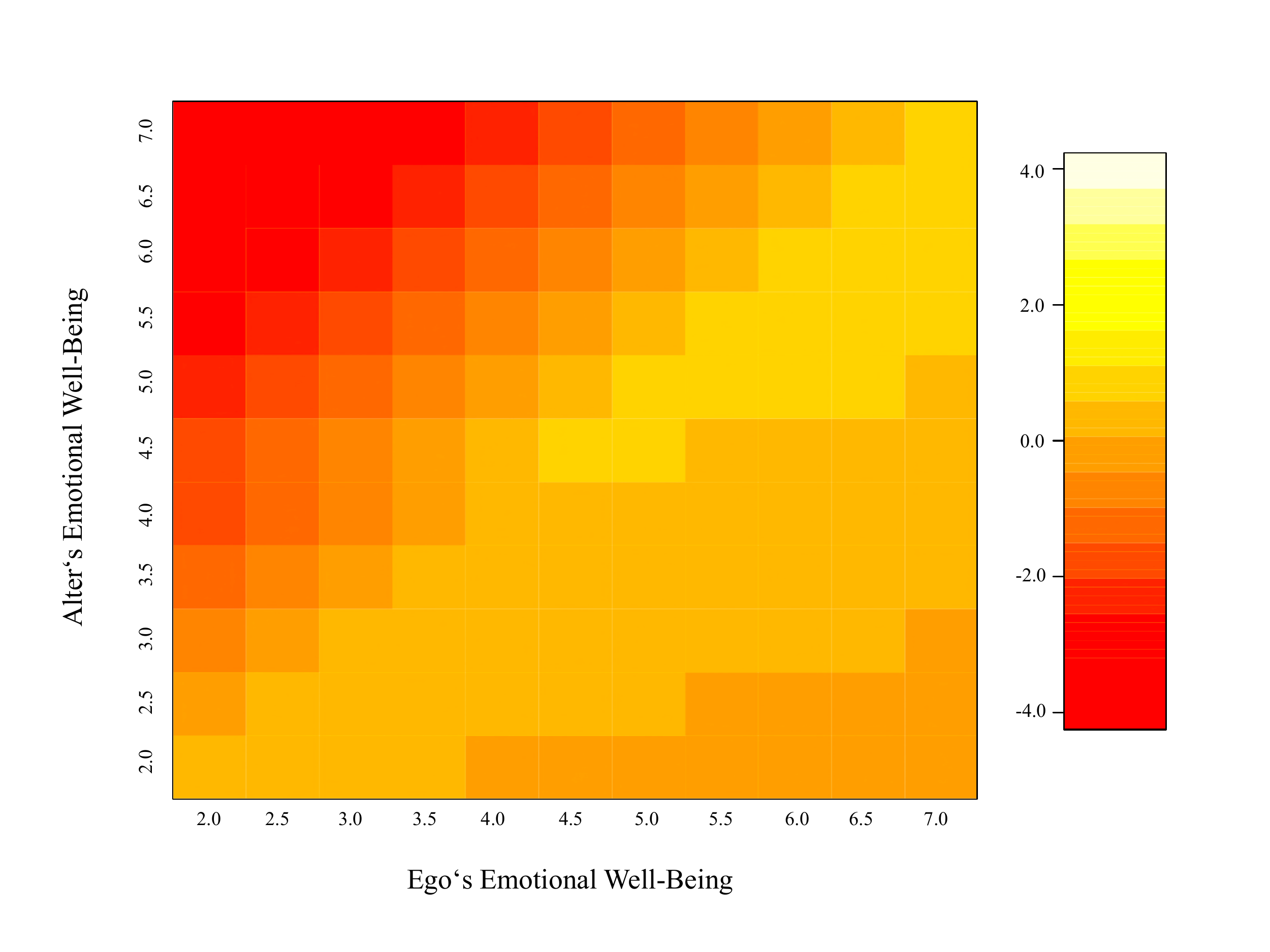}
    \caption[Ego-alter selection table]
      {\red{Heatplot of the ego-alter selection table. Only estimates for observed emotional well-being values (ranging from 2.0 to 7.0) are displayed. The colors of the cells indicate the relative gain to ego's objective function when choosing alters with various levels of emotional well-being (red = negative, yellow/white = positive).}}
    \label{selection}
\end{figure}
  \renewcommand{\baselinestretch}{2}
  
\red{Figure~\ref{selection} shows a heatmap of these values for each level of ego's and alter's emotional well-being in the strong friendship network. The columns correspond to ego's emotional well-being and the rows to alter’s emotional well-being. The colors of the cells indicate the relative gain to ego's objective function when choosing alters with various levels of emotional well-being (red = negative, green = positive). 
Figure~\ref{selection} shows that individuals with low emotional well-being tend to avoid nominating individuals with high emotional well-being in the strong friendship network.  Interestingly, this effect is much stronger than the tendency of individuals with high emotional well-being to avoid nominating alters with low emotional well-being. Also, the tendency to select individuals with similar levels of emotional well-being is higher between high emotional well-being individuals than between those with lower emotional well-being. 
}
\subsection{Covariates}
We examine whether any of the following covariates are related to emotional well-being: age as a continuous variable, and dummy variables for ethnic categories, religion, gender, and parental status. We conduct a score-type test for each variable with the final model. Results do not support that any of these covariates are significant predictors of emotional well-being ($\textit{p} > .05$).

\subsection{Differences between the evolution of weak and strong friendship networks}

We additionally observe a number of interesting differences when comparing the estimates of the weak and strong friendship networks in Table~\ref{SAOMs}. 
These differences emphasize the value of modeling the co-evolution of strong and weak networks with ordered SAOMs.
First, the negative transitive reciprocated triples effect is observed in the strong-tied network but not in the weakly tied network, indicating that individuals in the strong-tied network are less likely to reciprocate ties within triads.
Second, indegree-popularity and outdegree-activity effects contribute to the network evolution only in the weak-tied network but not in the strong-tied network. Third, becoming friends with friends of the partner is only observed in the weak-tied network whereas the closure of a couple four-cycle is only prevalent in the strong-tied network. Fourth, gender homophily, ethnic homophily, religion homophily, and age heterophily matter in the weak-tied network but not the strong-tied network.

\red{Finally, the \textit{strong mutuality with weak} cross-network effect is non-significant, which indicates that reciprocated ties of the weak-tied network do not predict ties of the same dyad in the strong-tied network ($\theta$ = -0.30, \textit{p} $>$ .10). This might suggest that weak ties are relatively stable and do not tend to transform into a strong tie if there is not a strong tie present already.}

\section{Discussion and conclusion}
This study investigates the social mechanisms underlying the co-evolution of weak and strong friendship ties and emotional well-being. It considers three groups of social mechanisms -- social integration,  social influence, and social selection -- which are argued to be relevant for individuals' well-being \citep{Berkman2000, VanZalk2010a}. The empirical setting of this study is a housing community for university affiliates and their partners. Most study participants moved to this community from other cities or countries within two years preceding the start of the study. Hence, how and with whom they become friends is expected to be important for their emotional well-being. %
The first contribution of this study is the theoretical discussion and empirical assessment of social integration mechanisms and emotional well-being. 
At the same time, we take social selection as an alternative mechanism into account to see whether individuals' emotional well-being explain the emergence of cohesive social structures. This way, we can gain more reliable results on the relationship of social ties and emotional well-being. %
Our analyses do not support the hypotheses that having many friends or dense social structures (i.e., social cohesion) lead to higher emotional well-being. We, however, find evidence that individuals with high emotional well-being nominate more friends in their strong-tied network. %

\red{
Former studies have demonstrated that extensive and cohesive friendship structures positively affect well-being. There are several potential reasons for the null findings regarding our social integration hypotheses. (1) Most network studies on social integration are conducted on adolescent populations \citep[e.g.,][]{Ueno2005}\nocite{Ueno2005}. Our sample consists of adults that moved into the community with their partners (and children), which indicates a more stable life situation. Furthermore, these individuals might already have an established set of friendships outside of this community that can be activated when needed. (2) Each individual in the sample already has one very close tie – their partner. As partners are one of the main sources of social support and well-being \citep{Dush2005, umberson1996}, the additional effect of being socially integrated in the community might be lower. (3) As suggested by Ueno (\citeyear{Ueno2005}), the relationship between social integration and mental health might be curvilinear, and this could have led to a linear null-finding. Unfortunately, with our method of analysis, we could not test this curvilinear hypothesis. (4) }Previous cross-sectional studies reporting positive effects of network size \citep[e.g.,][]{Schaefer1981a}\nocite{Schaefer1981a} and social cohesion \citep[e.g.,][]{Burt1987}\nocite{Burt1987} might be biased by not taking into account selection effect as we did in our study (since their cross-sectional study designs did now allow to differentiate between these effects).
\red{We indeed find strong evidence for the findings of Schaefer and colleagues (\citeyear{Schaefer2011}) that individuals with lower well-being (or higher depressive symptoms) tend to withdraw from their social network rather than being avoided by others. More specifically, we find that this effect appears because individuals with low emotional well-being tend to avoid nominating strong friends who report high levels of emotional well-being. Interestingly, we report a contrary effect for the weak-tied network, where there is a borderline significant tendency of individuals to prefer nominating others with high values of emotional well-being. Coyne's (\citeyear{Coyne1976}) prediction of avoiding alters with more depressive symptoms thus seems to be more supported for weak-tied friendship than strong-tied networks.} 

The results of this study unveiled a potential confounding mechanism leading to \red{ the commonly found association between the (reported) number of friends and well-being \citep{Brissette2000}.   %
It seems that this association might come into existence by withdrawal selection effects: individuals with lower emotional well-being report fewer strong-tied friends, especially to individuals with high levels of emotional well-being. This finding can be explained by various social mechanisms. First, individuals low in emotional well-being perceive strong friendships as less strong than their alters, which could be caused by negative social perceptions associated with depressive symptoms \citep{Gadassi2015, Gotlib1983}. Second, high well-being individuals are less liked by low well-being individuals because of upwards-comparison processes \citep{Festinger1954, Swallow1988}. Third, when individuals with low emotional well-being withdraw from their strong-tied network, they only find friends in marginal network positions whose well-beings are similarly low, which leads to the observed well-being homophily (as argued by Schaefer et al., 2011).} With our current study design, it is not possible to decide between these potential explanations, which thus invites further research.

The second contribution of this study is the investigation of social influence effects on emotional well-being while also taking the alternative mechanism of homophily into account \red{\citep{Mcpherson2001}}. We argue that individuals' emotional well-being is influenced by their strong-tied friends' emotional well-being and that two individuals with similar levels of emotional well-being are more likely to form or maintain strong ties (homophily).  %
The results do not provide conclusive evidence \red{for social influence mechanisms, as we only find borderline significant parameters for social influence through strong friendship ties in a simplified model. At the same time, our results indicate that emotional well-being homophily is a relevant mechanism in the evolution of the strong-tied friendship network. Furthermore, we report that emotional well-being homophily in a strong-tied friendship network is mostly driven by homophilic selection tendencies of individuals with high levels of emotional well-being. This is reasonable considering that being friends with someone of high well-being is generally more rewarding \citep{Lyubomirsky2006, Nezlek1994}.  As a consequence, individuals are more likely to form and maintain these friendship relations. This mechanisms will induce an over-representation of relationships of and between individuals with high emotional well-being.
} 

\red{We cannot rule out that emotional well-being homophily was the result of shared social foci that individuals with certain levels of emotional well-being engage in (i.e., secondary homophily, Shalizi \& Thomas, 2011)\nocite{Shalizi2010}. For example, individuals with higher emotional well-being might be more likely to attend social events, where they meet other individuals who also have high emotional well-being. 
}
\red{Furthermore, we did not consider that individuals in cohesive network positions might be more susceptible to social influence effects \citep{Guan2015a}.} %

Beyond these specific results, this study contributes to the research on social influence in several ways. %
First, it contributes to affect-based influence studies \citep[e.g.,][]{Joiner1999, Eisenberg2013}\nocite{Joiner1999, Eisenberg2013} by modeling the influence of multiple sources (i.e., alters with different affective states). 
Second, it contributes to longitudinal research in social networks on mental health \citep[e.g.,][]{VanZalk2010a}\nocite{VanZalk2010a} by introducing a purely affect-based measure and theoretical expectations about the influence of different qualities of friendships.   %

The third contribution of this study is to combine longitudinal network data with frequent measures of affect in one dynamic model. %
In this study, emotional well-being was measured through a cell phone-based experience sampling method (i.e., multiple assessments in real-life situations) which has several advantages over traditional data collection procedures. First, experience sampling methods minimize recency bias (i.e., that recent events which potentially affect well-being are weighted more). Second, individuals' well-being is assessed in their normal daily environment, which contributes to the ecological validity of this study \citep{Csikszentmihalyi1987, Reis2000}. Third, assessing emotional well-being through aggregated ratings of daily affect captures a distinct emotional quality of subjective well-being \citep{Diener2008}. At the same time, aggregating data from experience-sampling methods might suffer more from acquiescence biases (some individuals generally use lower or higher ends of scales) than traditional questionnaire methods \citep{Watson2002}.  %

The fourth contribution of this study is the application of \textit{ordered} Stochastic Actor-Oriented Models \citep[ordered SAOMs; ][]{Snijders2010, Ripley}, which allows us to simultaneously investigate how mechanisms of social integration, social influence, and social selection operate on weak and strong friendship relations. This study is among the first attempts to empirically test hypotheses in strong and weak networks.
Using ordered SAOMs helped us to get some interesting insights into co-evolution mechanisms between weak and strong friendship networks. For example, it seems that the creation and maintenance of weak ties is more affected by degree related effects (indegree-popularity, outdegree-activity), gender homophily, ethnic homophily, religion homophily, and age heterophily, whereas the tendency not to reciprocate ties within triads is only prevalent in the strong-tied network. Future research can build upon these initial findings and investigate co-evolutionary mechanisms between strong and weak ties in more detail. %
This study also has a number of limitations. 
First, the lack of social integration effects among our findings might be related to limitations in the empirical setting of the study. That is, our data only captured social integration within the given community. We did not investigate relationships outside of the community, therefore could not capture other social contexts in which individuals could achieve social integration. While we argue that this community should be relevant for the participants as it provides them with one of their very few local social contexts they have, several important longer-term relationships will most likely exist outside of this community. Hence, our theoretical arguments would benefit from testing these in additional distinct populations in the future. %
Second, we cannot be sure that the decision of individuals to omit the daily mobile questions or the online friendship questionnaire was not related to their current experience of emotional well-being or their network position even though our analyses on explaining missing values did not reveal consistent patterns. %
\red{Third, our measure for social cohesion was strictly local. It only captured individuals' local embeddedness within dense triads. Social cohesion within larger cliques of the network or not necessarily dense triads might be more related to well-being as the sense of belonging to a clique offers various benefits for an individual's well-being (e.g., social support, social identity; Thoits, 2011\nocite{Thoits2011}). For instance, Guan and Kamo (\citeyear{Guan2015a}) investigated influence of depressive symptoms between adolescents with macro-levels measures of network cohesion and report that those individuals in denser social structures are more susceptible to social influence.}
\red{The fourth limitation of this study concerns the measurement of emotional well-being. We measured emotional well-being with a one dimensional scale where positive and negative affect constituted the ends of the scale. However, a large body of research suggests that positive and negative affect are partially independent phenomena that result in distinct social behavior \citep{Berry1996a}. Further studies should differentiate between positive and negative dimensions of emotional well-being when investigating co-evolution of emotional well-being with social ties.}
Fifth, our study design might suffer from low statistical power. Even though we have a reasonably high number of friendship observations over the waves ($N_{weak} = 4'416, N_{strong} = 1'008$), the number of emotional well-being observations is limited ($N$ = 274). This scarcity of individual observations might be related to a number of non-findings.
One methodological conclusion of this study is thus a call for larger research designs (e.g., with more participants, more data collection waves, more daily assessments) to investigate the presented theoretical mechanisms about changes in emotional well-being \citep{Stadtfeld2016a}. Such larger research designs should further investigate gender-specific effects that are associated with social influence \citep[e.g.,][]{VanZalk2010b}\nocite{VanZalk2010b}, social integration \citep[e.g.,][]{Falci2009}\nocite{Falci2009} and social selection \citep[e.g.,][]{Block2014}\nocite{Block2014}.
Our study makes important contributions to the understanding of social mechanisms that explain how individuals experience emotional well-being and how emotional well-being explains the emergence of social networks. We argue that weak and strong friendship ties co-evolve with emotional well-being and offer a novel methodological approach (ordered SAOMs) to test hypotheses on social integration, social influence and social selection. We find support for the prevalence of social influence and social selection. 
The theoretical model introduced in this study calls for future empirical research investigating social mechanisms that sustain or improve aspects of subjective well-being. This study takes an important step towards Thoits' (\citeyear{Thoits2011}) instigation for research on \textit{how} social ties matter for mental health and well-being. It is the first study to simultaneously investigate social integration and social influence processes that are both argued to play a major role in the manifestation of well-being \citep{Berkman2000}. At the same time it takes into account processes of social selection and distinguishes effects related to strong and weak social ties. These steps towards the understanding of social mechanisms underlying subjective well-being are highly relevant considering the crucial importance of well-being for individuals and societies. %

\newpage

\linespread{1}
\scriptsize
\bibliography{/Applications/TeX/library}

\newpage

\section{Appendix}
\renewcommand{\baselinestretch}{1}
\begin{table}[ht]
  \caption{Parameter direction and significance levels of emotional well-being effects for alternative model specifications}
  \label{robust}
  \begin{minipage}{\textwidth}
\begin{tabular}{
L{6cm}
L{1cm}
L{1cm}
L{1cm}
L{1cm}
L{1cm}
L{1cm}
L{1cm}
}
\hhline{========}
 & \multicolumn{3}{c}{Weak}  &  \multicolumn{3}{c}{Strong} \\
 \hline
 & Alter & Ego & Sim & Alter & Ego & Sim & avSim \\
\hline
Original model & + $\dagger$ & 0 & 0 & - * & + ** & + * & 0 \\
EWB rounded to full numbers & + * & 0 & 0 & - * & + * & + * & 0 \\
EWB median & + $\dagger$ & 0 & 0 & 0 & + * & 0& 0 \\
Weak = 3 & + ** & 0 & - * & 0 & + * & 0& 0 \\
Strong = 6 & + $\dagger$ & 0 & 0 & 0 & + * & 0& 0 \\
EWB aggregated over  20 days & + *** & 0 & 0 & 0 & + * & 0 & 0\\
EWB aggregated over  40 days & + ** & 0 & - * & 0 & + * & 0 & 0\\
\hhline{========}
\end{tabular}
 \end{minipage}
 \newline \\
\begin{tablenotes}
      \item \textit{Note.} EWB = emotional well-being, \red{$+$ = positive parameter estimate, $-$ = negative parameter estimate, 0 = non-significant effect, $\dagger$ \textit{p} $<$ .10, * \textit{p} $<$ .05, ** \textit{p} $<$ .01, *** \textit{p} $<$ .001. Two-sided \textit{p}-values. Parameters estimated with Methods of Moment.}
    \end{tablenotes}
\end{table}

\begin{figure}
    \includegraphics[scale=0.7]{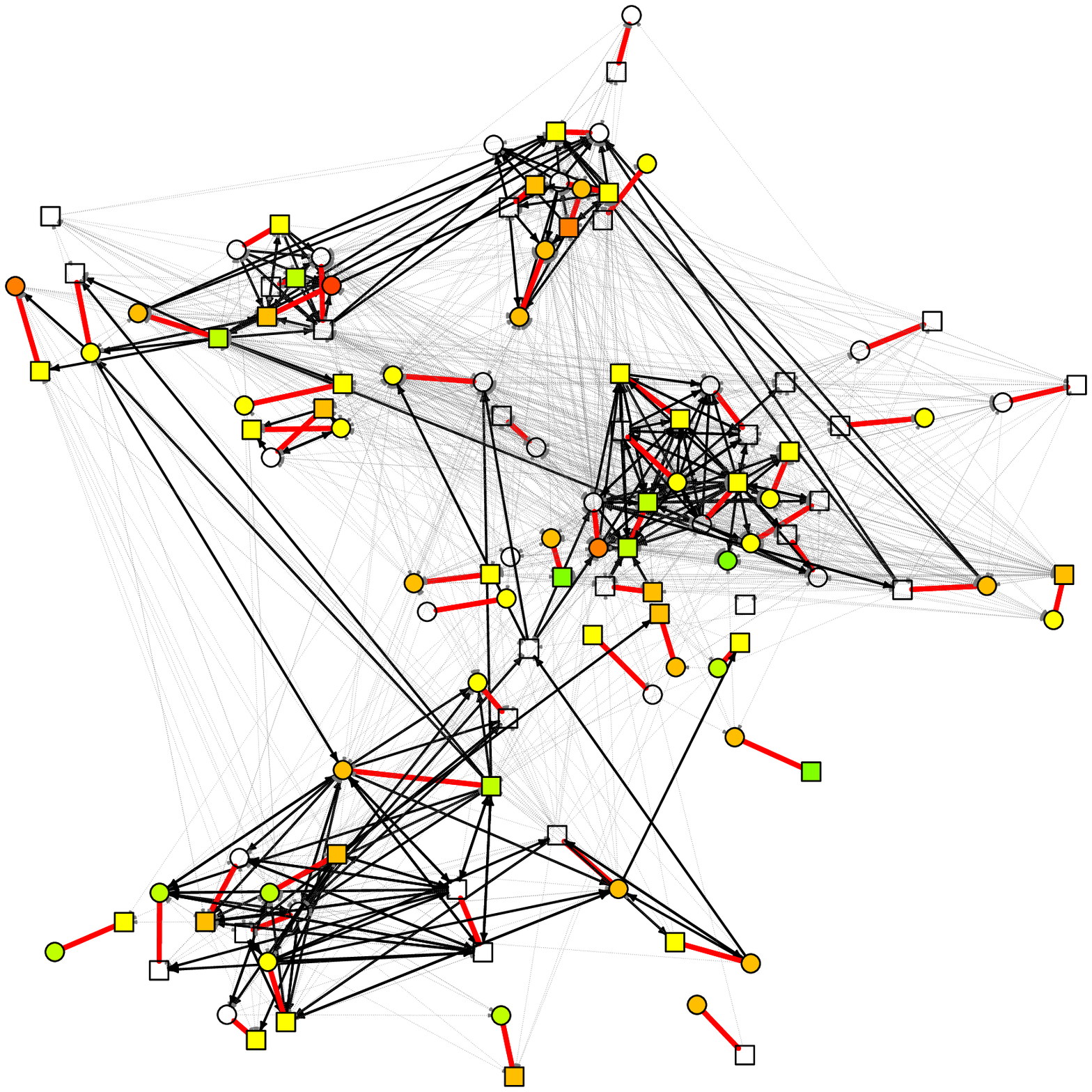}
    \caption[\red{Friendship network at T3 with weak, strong, and partnership ties}]
      {Friendship network at wave three. Partnership ties are marked red, weak ties with thin dotted lines, and strong ties as solid lines. Continuous node color represents well-being scores (red = low score, green = high score, white = missing). \red{Squared node shapes are male participants, whereas round node shapes are female.}}
    \label{netplot2}
\end{figure}

\end{document}